\documentclass[11pt]{article}
\usepackage{latexsym}
\usepackage{amssymb}
\usepackage{epsf}
\usepackage{amsmath}
\usepackage{graphicx}% Include figure files
\usepackage{slashed}
\usepackage{color}

\begin{document}
\pagestyle{empty}

\begin{flushright}
KEK-TH-1932\\
MAD-TH-16-06
\end{flushright}

\vspace{3cm}

\begin{center}

{\bf\LARGE Primordial Lepton Oscillations and Baryogenesis} 
\\

\vspace*{1.5cm}
{\large 
Yuta Hamada$^{1,2}$ and Ryuichiro Kitano$^{1,3}$
} \\
\vspace*{0.5cm}

{\it
$^1$KEK Theory Center, Tsukuba 305-0801, Japan\\
$^2$Department of Physics, University of Wisconsin, Madison, WI 53706, USA\\
$^3$Department of Particle and Nuclear Physics\\
The Graduate University for Advanced Studies (Sokendai)\\
Tsukuba 305-0801, Japan\\
}

\end{center}

\vspace*{1.0cm}

\begin{abstract}
{\normalsize 
The baryon asymmetry of the Universe should have been produced after the
 inflation era. We consider the possibility that the asymmetry is
 generated by the flavor oscillations in the reheating process after
 inflation, so that the baryon asymmetry is realized already at the
 beginning of the radiation dominated era. In the seesaw model, we show
 that the propagators of the left-handed leptons generically have flavor
 mixings in the thermal background, that can generate flavor-dependent
 lepton asymmetry through the $CP$ violation in the oscillation
 phenomena. The flavor dependent rates for the wash-out process can
 leave the net asymmetry today.  }
\end{abstract} 

%%%%%%%%%%%%%%%%%%%%%%%%%%%%%%%%%%%%%%%%%%%%%%%%%%%%%%%%%%%%%%%%%%%%%%%%%%%%
\newpage
\baselineskip=18pt
\setcounter{page}{2}
\pagestyle{plain}
\baselineskip=18pt
\pagestyle{plain}

\setcounter{footnote}{0}

%%%%%%%%%%%%%%%%%%%%%%%%%%%%%%%%%%%%%%%%%%%%%%%%%%%%%%%%%%%%%%%%%%%%%%%%%%%% 
% Introduction 
%%%%%%%%%%%%%%%%%%%%%%%%%%%%%%%%%%%%%%%%%%%%%%%%%%%%%%%%%%%%%%%%%%%%%%%%%%%% 
\section{Introduction}

The $CP$ violation in the neutrino oscillation can produce flavor
dependent particle-antiparticle asymmetry. Although it has not been
established experimentally, the discovery of such phenomena will be a
quite important hint for the mystery of the baryon asymmetry of the
Universe. The mechanism for the baryon asymmetry before the electroweak
phase transition requires generation of primordial $B-L$ asymmetry,
rather than $B$ asymmetry, due to the $B+L$ breaking electroweak
sphaleron process~\cite{Kuzmin:1985mm}. Within the field content of the
Standard Model, the Majorana neutrino mass term, $ll HH$, is the lowest
dimensional operator which breaks $B-L$ explicitly. Therefore, it is
quite reasonable that the $CP$ violation in the neutrino interactions is
responsible for baryogenesis.

The standard leptogenesis scenario~\cite{Fukugita:1986hr} assumes that
the $CP$ violation in the decays of right-handed neutrinos produces the
lepton asymmetry, and thus it requires the production of the
right-handed neutrinos after the inflation and their decays when the
$L$-violating interactions gets sufficiently ineffective. It has been
extensively studied such possibilities and viable parameter regions are
investigated. It has been found that the mechanism works for high
enough reheating temperatures. For the review, see
Ref.~\cite{Buchmuller:2005eh} for example.

There is another interesting period of time where the out-of-equilibrium
condition is satisfied automatically. After the inflation, the decays of
the inflaton can reheat the Universe, producing the radiation energy
which eventually dominates the Universe. In the course of reheating,
high-energy particles are produced by the decay process, and each
particles lose their energy by scattering with thermal plasma. This
energy-loss process is obviously a one-way process, and thus provides us
with perfect environment for baryogenesis. In
Ref.~\cite{Hamada:2015xva}, such a possibility has been studied where
the scattering of the leptons during the reheating process produces the
baryon asymmetry of the Universe.

In this paper, we consider the oscillation phenomena of the left-handed
leptons (neutrinos and charged leptons) in the thermal background and
the possibility to produce lepton asymmetry via its $CP$ violation.\footnote{
See Ref.~\cite{Garbrecht:2012pq} for baryogenesis from the oscillations of left-handed leptons which are originated from the decay of right-handed neutrinos.}
The inflaton decays into leptons in a particular combination of the flavor
eigenstates. Since the lepton propagators in the thermal background are
not diagonal in the flavor basis, the leptons can change the flavor
during the propagation. The $CP$ violation in the oscillation, i.e., the
indirect $CP$ violation in the inflaton decays, can produce the flavor
dependent lepton asymmetry although no net lepton asymmetry is
generated. The flavor dependent lepton asymmetry, in turn, can be
converted into the net asymmetry by the flavor dependent wash-out
process due to, for example, the $llHH$ interactions. We find that the
baryon asymmetry of the Universe can be explained by this mechanism,
even in the case where the reheating temperature is much lower than the
masses of the right-handed neutrinos and there is no decays of the
inflatons into right-handed neutrinos, which, therefore, have never
shown up in the history of the Universe.

This paper is organized as follows. 
In Sec.~\ref{sec:propagator}, we discuss the lepton oscillation
phenomena in the context of the seesaw model.
We show that observed baryon asymmetry of the Universe can be explained
by the primordial lepton oscillation in Sec.~\ref{sec:asymmetry}.
Sec.~\ref{sec:conclusion} is devoted for summary and conclusion.

\section{Lepton propagators in the early Universe}\label{sec:propagator}

In this section, we first calculate the propagators of the left-handed
leptons in the thermal background, and show that the thermal correction
to the dispersion relation causes the flavor oscillations as in the case
of the neutrino oscillation in the vacuum by the mass differences.
%Then, the $CP$ violation in the lepton oscillation is evaluated.
%We first calculate the propagators of the left-handed leptons in the
%thermal background. 
As a concrete scenario, we consider the seesaw model~\cite{typeI} in which the
left-handed leptons have flavor off-diagonal interactions.

The Lagrangian is given by
\begin{align}
 {\cal L}_{\rm int} = 
& - y_\nu^{ij} \bar N_i P_L (l_j \cdot \tilde H) +
 {\rm h.c.}
\nonumber \\
& + {M_i \over 2} \bar N_i^{c} N_i   +
{\rm h.c.}
%& + {M_i \over 2} \bar N_i N_i
\nonumber \\
& + y_e^{l} \bar e_l P_L (l_l \cdot H) + {\rm h.c.}
\label{eq:lagrangian}
\end{align}
We have diagonalized the charged lepton Yukawa coupling $y_e$, that defines
the flavor eigenstate. The right-handed neutrinos, $N_i$, are introduced
and $M_i$ are their Majorana masses.

%\begin{align}
% m_\nu^{ij}& = y_\nu^{ki} M_k^{-1} y_\nu^{kj} \langle H \rangle^2.
%\end{align}
%\begin{align}
% U_{\rm MNS}^{ik} m_\nu^{ij} U_{\rm MNS}^{jl} = m_\nu^{k} \delta_{kl}
%\end{align}
%\begin{align}
% y_\nu^{ij}& = \sqrt{M_i} R_{ik} \sqrt{m_\nu^k} (U_{\rm MNS}^{jk})^* /
% \langle H \rangle.
%\end{align}

%Story: 
%\begin{itemize}
% \item Inflaton decays preferably to third generation particles.
% \item A part of tau leptons are converted to electrons and muons
%       through the neutrino oscillation phenemena induced by the flavor
%       non-diagonal thermal masses.
% \item The CP violation in the oscillation generates the flavor
%       dependent lepton number asymmetry.
% \item The flavor dependent lepton number violating interaction
%       partially erase the asymmetry with the flavor depenent rates.
% \item The net lepton asymmetry remains and converted to the baryon asymmetry.
%\end{itemize} 

%\subsection{Medium effects for neutrino propagation}

In a high temperature medium where the Higgs particles are in the
thermal bath, the Yukawa interactions in Eq.~\eqref{eq:lagrangian}
affect the propagators of the $l_i$ fields.
Following Ref.~\cite{Weldon:1982bn}, the propagator of a fermion field
in the momentum space is parametrized as
\begin{align}
 S(K)& = \left[
(1+a) \slashed K + b \slashed u
\right]^{-1},
\end{align}
where $K^\mu = (\omega, \bf k)$ is the four momentum and $u^\mu =
(1,0,0,0)$. The coefficients, $a$ and $b$, are functions of $\omega$ and
$k = | \bf k |$. By looking at the pole of the propagator, the
dispersion relation is modified to
\begin{align}
 \omega & = k - {\rm Re} \left[ {b \over 1 + a} \right].
\label{eq:dispersion}
\end{align}
The non-vanishing functions, $a$ and $b$, are the effects of the
interactions.  At the leading order in the perturbation theory, one can
ignore $a$. The real part of the function $b$ is calculated to be
\begin{align}
 b_{ll'}& \simeq - {T^2 \over 16 k} (y_e^l)^2 \delta_{ll'}
+ {\pi^2 T^4 \over 9 M_k^4} y_\nu^{kl*} y_\nu^{kl'} k,
\label{eq:b-term}
\end{align}
for $\omega \sim k \gg T$, where $T$ is the temperature.
We have assumed here that $Tk \ll M_i^2$. For $Tk \gtrsim M_i^2$, one obtains the term
similar to the first term which stems from the charged lepton Yukawa
interactions.
The second term provides flavor non-diagonal entries, responsible for
the oscillation. 
The second term is not quite the ``thermal mass,'' since the dispersion
relation is still $\omega = 0$ in the $k \to 0$ limit. However, it does
modify the dispersion relation as in Eq.~\eqref{eq:dispersion}, and
causes the flavor oscillation phenomena as we discuss below.
The contribution from the gauge interactions are flavor universal, and
thus can be ignored for our purpose. 

The second term is suppressed by $M_i^4$, which can be understood by the
operator analyses. For $M_i \gg T$, one can integrate out the $N_i$
fields, and effective interaction terms with dimension five or higher
are generated. The odd-dimensional operators, such as $llHH$ and $ll H
\partial^2 H$, break the $L$ number as well as the Higgs number, and
thus would not cause the forward elastic scattering to modify the
propagators. The dimension-six operators, $\bar l \gamma^\mu l (i
H^\dagger \partial_\mu H + {\rm h.c.})$, would not contribute as long as
there is no chemical potential for the Higgs fields. Therefore, the
first contribution appears from the dimension-eight operators which are
suppressed by $M_i^4$.
Nevertheless, for a large enough $k$,
\begin{align}\label{eq:large k condition}
k^2\geq
{9 \over16\pi^2} {(y_e^l)^2 \over y_\nu^{kl*} y_\nu^{kl'}} {M_k^2 \over T^2} M_k^2,
\end{align}
the second term in Eq.~\eqref{eq:b-term} dominates over the first one.
%

%By integrating out the right-handed neutrinos, we obtain a term like:
%\begin{align}
% {\cal L}_{\rm eff}& = M_i^{-2}
%y_\nu^{ij*} y_\nu^{ik}
%\left(
%\bar l_j \gamma^\mu l_k
%\right)
% (i\tilde H^\dagger \partial_\mu \tilde H
% - i\partial_\mu \tilde H^\dagger  \tilde H ).
%\end{align}
%In a thermal background, this term cause the appearance of the effective
%Hamiltonian:
%\begin{align}
% H_{\rm eff}& \sim M_i^{-2}
%y_\nu^{ij*} y_\nu^{ik} \times {T^3} \equiv m_{\rm eff}^{jk}.
%\end{align}

The matrix $b$ can be diagonalized
by a unitary transformation:
\begin{align}\label{eq:diagonal}
 U^\dagger b U = b^{\rm diag}.
\end{align}
The unitary matrix $U$ is different from the PMNS matrix~\cite{PMNS}, and the actual form cannot
be determined only from the low energy data. 
%
%
%\subsection{neutrino oscillation}
%
%The field theoretic computation involves the following factor in the
%transition rate from the neutrino production by the inflaton decay to
%the first scattering which identifies the flavor state:
%\begin{align}
%& {1 \over \pi} \int_0^\infty d \omega
%{1 \over (\omega + b_i)^2 - k^2 + i \omega \Gamma}
%{1 \over (\omega + b_j)^2 - k^2 - i \omega \Gamma}
%\nonumber \\
%& \simeq {1 \over \pi} \int_0^\infty d \omega
%\left[
%- i \pi \delta \left(
%(\omega + b_i)^2 - k^2
%\right) {1 \over (\omega + b_j)^2 - k^2 - i \omega \Gamma}
%%
%+ i \pi \delta \left(
%(\omega + b_j)^2 - k^2
%\right) {1 \over (\omega + b_i)^2 - k^2 + i \omega \Gamma}
%\right]
%\nonumber \\
%& = {1 \over \pi} {1 \over 2 k} (-i \pi)
%\left[
%{1 \over (k - b_i + b_j)^2 - k^2 - i (k - b_i) \Gamma}
%%
%+ {1 \over (k - b_j + b_i)^2 - k^2 + i (k - b_j) \Gamma}
%\right]
%\nonumber \\
%& \simeq - {i \over 2 k} 
%\left[
%{1 \over - 2 (b_i - b_j) k  - i k \Gamma}
%+ 
%{1 \over 2 (b_i - b_j) k + i k \Gamma}
%\right]
%\nonumber \\
%& = {i \over k^2} 
%\left[
%{1 \over - 2 \Delta b_{ij}  + i \Gamma}
%\right]
%\nonumber \\
%& = {1 \over k^2 \Gamma} 
%\left[
%{1 \over 1 + 2 i \Delta b_{ij} / \Gamma }
%\right].
%\end{align}
%The imaginary part of it is given by
%\begin{align}
%{\rm Im} \left[
%{1 \over 1 + 2 i \Delta b_{ij} / \Gamma }
%\right]
%= { - 2 \Delta b_{ij} / \Gamma
%\over
% 1 + (2 \Delta b_{ij} / \Gamma )^2 }.
%\end{align}
%This factor causes the $CP$ violation in the neutrino (and charged
%lepton) propagation.
%
The neutrino oscillation can be understood in the standard quantum
mechanical considerations. In the basis of diagonal $b$, the neutrino
wave function is given by~\cite{PDG}
\begin{align}
 | t = \Delta t \rangle& = e^{- i \omega_i \Delta t} e^{i k_i \Delta x} | t = 0 \rangle,
\end{align}
and the differences of the dispersion relations appear in the phase of the
interference terms in the transition rates of $l_l\to l_{l'}$:
\begin{align}\label{eq:QM}
 \delta \phi_{jk}& = (\omega_j - \omega_k) \Delta t - (k_j - k_k) \Delta x
\nonumber \\
& = (k_j - k_k) (\Delta t- \Delta x) - \Delta b_{jk} \Delta t
\nonumber \\
& \simeq - \Delta b_{jk} \Delta t,
\end{align}
where $\Delta b_{jk} = b_j^{\rm diag} - b_k^{\rm diag}$, and $\Delta t$
and $\Delta x$ are time and length of the travel, respectively. The time
$\Delta t$ should be taken as the mean free time of the lepton
propagation, $\Delta t \simeq \Gamma^{-1}$, after which the leptons lose
their momenta or pair annihilate with the leptons in the thermal
bath. The dominant scattering process is through the $SU(2)$ gauge
interactions which give a flavor universal $\Gamma$:
\begin{align}
  \Gamma \sim {g_2^2 \over 4 \pi} T.
\label{eq:gamma}
\end{align}
%
%\textcolor{red}{Notice that, in Eq.~\eqref{eq:QM}, if we take $\omega_j=\omega_k$, $k_j=k_k$, $T=L$ all three choices lead same conclusion.}
%, that reproduce the oscillation factor in the field
%theoretic computations.

Essentially the same result can be derived from field
theory~\cite{ArkaniHamed:1997km}. 
In the basis of diagonal $b$, the denominator of the propagators is
\begin{align}
%{1 \over D} = 
{1 \over (\omega + b_i^{\rm diag})^2 - k^2 + i \omega \Gamma },
\end{align}
where $\Gamma$ is given in Eq.~\eqref{eq:gamma}.
%where the width $\Gamma$ is given by the flavor universal gauge
%interactions,
%\begin{align}
% \Gamma \sim {g_2^2 \over 4 \pi} T.
%\end{align}
%Here $g_2$ is the $SU(2)$ gauge coupling.
%
The field theoretic computation of the
probability involves the phase space integration over the propagators:
\begin{align}
& \int_0^\infty d \omega
{1 \over (\omega + b_j)^2 - k^2 + i \omega \Gamma}
{1 \over (\omega + b_k)^2 - k^2 - i \omega \Gamma}
 \simeq {\pi \over k^2 \Gamma} 
\left[
{1 \over 1 + 2 i \Delta b_{jk} / \Gamma }
\right].
\end{align}
We see the oscillation effect $1 / (1 + 2 i\Delta b_{jk} / \Gamma)$, which is 
similar to the quantum mechanical consideration, $\exp (- i\Delta b_{jk}
/\Gamma)$, for $\Delta b_{jk} / \Gamma \ll 1$.

\section{Baryon asymmetry}\label{sec:asymmetry}

Now we consider the possible scenario for realizing the baryon asymmetry
through the primordial oscillation phenomena.

We assume that the high energy leptons are generated by the decays of
the inflaton $\phi$:
\begin{align}\label{eq:decay}
 \phi& \to l_i + X
\end{align}
and its $CP$ conjugate process. The part of the final state $X$ is
arbitrary, and the decay mode above is not even necessary to be the
dominant one. For example, $X = H + e^c_j$ makes $\phi$ gauge singlet. 
We also assume that the reheating temperature of the Universe, $T_R$, is lower
than $10^{12}$~GeV so that the Yukawa interaction of the tau leptons
is in the thermal equilibrium~\cite{Cline:1993bd}.
In this circumstance, one can distinguish lepton asymmetries in $l_\tau$
and $l_{e,\mu}$ as in the case of flavored
leptogenesis~\cite{Nardi:2006fx}, where indices are defined in the basis
where the charged lepton Yukawa matrix is diagonal.

As we have seen in the previous section, the neutrino Yukawa interaction
is not diagonal in this basis. The leptons obtain the off-diagonal
components in their propagator in the thermal plasma as in
Eq.~\eqref{eq:b-term}. Therefore, the leptons generated through the
inflaton decays undergo the flavor oscillations until the first
scattering where the leptons lose their energy.

In the basis where the propagator is diagonalized, the flavor eigenstate
is expressed as
\begin{align}\label{eq:final}
 | l ( e, \mu, \tau ) \rangle & = \sum_{j=1,2,3} U^\dagger_{jl} | j \rangle,
\end{align}
and the inflaton decays provide lepton states which are a
linear combination of the eigenstates of the Hamiltonian:
\begin{align}\label{eq:initial}
 | l_{\phi} \rangle & = \sum_{j=1,2,3} V^*_{j} | j  \rangle.
\end{align}
Here, $U$ is a unitary matrix same as that in
Eq.~\eqref{eq:diagonal}, whereas $V$ is a normalized vector.  By
combining Eqs.~\eqref{eq:QM}, \eqref{eq:final}, and \eqref{eq:initial},
the $CP$ asymmetry is given by the oscillation
formula~\cite{PDG,Krastev:1988yu}:
\begin{align}
 P_{l} - P_{\bar l}& = 4 \sum_{j>k} {\rm Im} \left(
U_{lj} V_{j}^* V_k U^*_{lk} \right)
\int dk f(k)
\sin {\Delta b_{jk}(k) \over \Gamma}.
\end{align}
Here $P_{l(\bar l)}$ represents the oscillation probability of
$l_\phi\to l_l\,(\bar l_\phi\to\bar l_l)$, 
$f(k)$ is the momentum distribution of neutrinos from inflaton decay with $\int dkf(k)=1$,
and we have evaluated
asymmetry at the stage of the first scattering, that is, $\Delta t\simeq
\Gamma^{-1}$.
In the following, we take $V_1 = 0$ for simplicity. 
Then, the $CP$ asymmetry is
\begin{align}\label{eq:oscillation formula}
 P_{l} - P_{\bar l}& = 4 
{\rm Im} \left( U_{l3} V_3^* V_2 U^*_{l2} \right)
\int dk f(k)
\sin {\Delta b_{32}(k) \over \Gamma}
\nonumber\\
&\simeq 4 
{\rm Im} \left( U_{l3} V_3^* V_2 U^*_{l2} \right)
{\Delta b_{32}(m_\phi) \over \Gamma}.
\end{align}
In the second line, we have expanded the sine function, and used the relation that the expectation value of momentum, $\int dkkf(k)$, is the order of $m_\phi$ which is the mass of the inflaton. %we have also assumed that the second term dominates in Eq.~\eqref{eq:b-term}. 
Eq.~\eqref{eq:oscillation formula} is generally non-vanishing and the $CP$-asymmetry factor, $A_{CP} =
{\rm Im} (U_{l3} V_3^* V_2 U^*_{l2})$, can easily be of order unity.
Since the asymmetry vanishes when we take the sum over the $l$ index,
there is no net lepton asymmetry generated. The tau asymmetry is
compensated by the $e$ and $\mu$ asymmetry.

At the stage of the first scattering, however, the flavor eigenstate is
not a meaningful concept since the time scale of the first scattering is
much faster than the scattering through the Yukawa interaction for the
charged tau lepton. This means that the density matrix of the lepton number
is not diagonal in the flavor basis. The flavor dependent asymmetry
becomes physical later at the time scale of the scattering through the
tau Yukawa interaction, where the off-diagonal components of the density
matrix dump to zero while the diagonal components, which we calculated,
are left time-independent~\cite{Nardi:2006fx}.
Note that, if the interaction rate of neutrino Yukawa couplings is
larger than that of tau Yukawa, it could generate the off-diagonal
components of the density matrix, which threatens our baryogenesis
mechanism.
As long as right-handed neutrinos are heavy, it is safe because we know
that the dominant interaction rate through the $llHH$ operator is
smaller than Hubble rate as we discuss later\footnote{We thank Sacha
Davidson for discussion on the flavor basis and the time scales of the
various interactions.}.

Since we assume that the temperature is low enough that the tau Yukawa
interaction is in the thermal equilibrium, one can identify the tau
asymmetry, and thus the generated flavor dependent asymmetry can be
separately treated. In particular, the $\Delta L = 2$ process through
the $ll HH$ interaction can wash out the lepton asymmetry in a flavor
dependent manner, that results in the generation of the net lepton
asymmetry.
If we assume the hierarchical neutrino masses in normal hierarchy, one
can consider the situation where only the $l_3 l_3 H H$ interaction is
effective. The interaction rate for the $\Delta L = 2$ process is
approximately given by
\begin{align}
 \Gamma_{l}^{\rm w.o.}& \simeq 
{|U^{\rm PMNS}_{l3}|^2 \over \pi^2 } {m_{\nu 3}^2 \over v^4} T_R^3,
\end{align}
where $U^{\rm PMNS}$ is the PMNS matrix, $m_{\nu i}$ is the neutrino
mass, and $v\simeq246$GeV.  Compared with the expansion rate of the
Universe, we obtain
\begin{align}
 {\Gamma_l^{\rm w.o.} \over H}&
= {T \over 10^{12}~{\rm GeV}}
 \times \left \{
\begin{array}{ll}
 0.03, & l = \tau \\
 0.02, & l = \mu + e \\
\end{array}
\right.
\end{align}
Here we take the observed values of mixing angles and
$m_{\nu3}\simeq0.05$~eV~\cite{PDG}.  Note that this quantity does not
depend on the Dirac or Majorana phases.

As a result,
we obtain the net lepton asymmetry as
\begin{align}
 {n_L \over s}& \sim {n_\phi \over s}
 \times B_{\phi \to l}
 \times A_{CP} \times {\Delta b_{23} \over
 \Gamma} \times 
\left(
 {\Gamma_\tau^{\rm w.o.} \over H}- {\Gamma_{\mu+e}^{\rm w.o.} \over H}
\right)
\nonumber\\
&
\sim {n_\phi \over s} 
\times B_{\phi \to l}
\times A_{CP} \times {\Delta b_{23} \over
 \Gamma} \times 0.01 \times 
\left(
{T_R \over 10^{12}~{\rm GeV}}
\right),
\end{align}
%}
where $B_{\phi \to l}$ is the branching ratio of the inflaton decays
into leptons.  
The first factor is the inflaton abundance, and the factor of the difference
of the $\Gamma^{\rm w.o.} / H$ describes the flavor-dependent wash-out effects.
The number density of the inflaton is
\begin{align}
 {n_\phi \over s} & \simeq {T_R \over m_\phi},
\end{align}
and the splitting of the dispersion relation is
\begin{align}
 { \Delta b_{23} \over \Gamma }&
\sim {4 \pi^3 m_\phi T_R^3 \over 9 g_2^2 M_k^4} |y_{\nu}^{k3}|^2,
\end{align}
by taking $k \sim m_\phi$. %where $m_\phi$ is the mass of the inflaton.
Therefore, putting altogether, we obtain the baryon asymmetry
\begin{align}
 {n_B \over s} \sim {n_L \over s}& \sim 10^{-7} \times 
B_{\phi \to l}
\times
A_{CP} 
\times
\left(
{T_R \over 10^{12}~{\rm GeV}}
\right)^2
\left(
{10 \over M / T_R}
\right)^{-3},
\end{align}
where $M$ is the right-handed neutrino mass which we take, for
simplicity, to be common for all flavors. We see that the baryon
asymmetry, $n_B / s \sim 10^{-10}$~\cite{Ade:2015xua}, can be explained
by this mechanism. It is interesting to note that the final formula is
independent of the inflaton mass as long as $m_\phi \gg T_R$.  We also
note that Eq.~\eqref{eq:large k condition},
\begin{align}
k\geq
{3\over4\pi} {y_e^l \over \sqrt{y_\nu^{kl*} y_\nu^{kl}}} {M\over T_R} M
\sim
10^{12}~{\rm GeV} 
\left({y_e^l\over10^{-2}}\right) \left({M\over10^{13}\text{GeV}}\right)^{3/2}
\sqrt{0.05\text{eV}\over m_{\nu3}}\left({10^{12}\text{GeV}\over T_R}\right),
\end{align}
can be satisfied for large $m_\phi\sim k$.

The estimate above is only valid when $M^2 \gg T_R m_\phi$. For more
general situations, the wash-out rates as well as the thermal correction
to the propagators can be enhanced, and thus a larger asymmetry may be
realized unless the neutrino Yukawa couplings are larger than tau Yukawa
coupling.  We leave such general analyses for future studies.

\section{Conclusion}\label{sec:conclusion}

In the inflationary Universe, one of the natural possibilities to
explain the baryon asymmetry is that the asymmetry is generated during
the reheating era such that the radiation dominated Universe has started
with non-vanishing asymmetry.
Since the transition from the inflaton dominated era to the radiation
dominated one is a one-way process, one can naturally satisfy the
out-of-equilibrium condition for baryogenesis. Also, for high enough
reheating temperatures, the $L$-violating interactions responsible for
the neutrino masses and the electroweak sphaleron process is effective,
that can provide the $B$-violating condition. The final condition is the
$CP$ violation. In the reheating era, the natural location to look for
the $CP$ violation is the decays of the inflaton. We indeed have shown
that $CP$ violation in the flavor oscillation of the leptons can happen
in the inflaton decays and that can explain the baryon asymmetry of the
Universe.

\vspace{1em}
\section*{Acknowledgements} 
We would like to thank Sacha Davidson for useful comments. This work is
supported by JSPS KAKENHI Grant-in-Aid for Scientific Research (B)
(No. 15H03669 [RK]), MEXT Grant-in-Aid for Scientific Research on
Innovative Areas (No. 25105011 [RK]) and Grant-in-Aid for JSPS Fellows
(No. 16J06151 [YH]).

%%%%%%%%%%%%%%%%%%%%%%%%%%%%%%%%%%%%%%%%%%%%%%%%%%%%%%%%%%%%%%%%%%%%%%%%%%%%%%%%%
% References 
%%%%%%%%%%%%%%%%%%%%%%%%%%%%%%%%%%%%%%%%%%%%%%%%%%%%%%%%%%%%%%%%%%%%%%%%%%%%%%%%%

\end{document}